\documentclass[pra,aps,twocolumn,superscriptaddress,showpacs]{revtex4}
\usepackage{graphicx}
\usepackage{amsmath}

\newcommand{\eq}{\begin{equation}}
\newcommand{\fine}{\end{equation}}

\begin{document}

\title{  \bf \Large  Generic Contractive States  and  Quantum  Monitoring  of\\
Free Masses and Oscillators.\footnote{Submitted to Physics News.} }

\date{March 20, 2019}

\author{Priyanshi Bhasin}
\affiliation{ 75/L, Model Town, Rewari, Haryana}
\author{Ujan Chakraborty}
\affiliation{ Indian Institute of Science Education And Research, Kolkata}
\author{ S . M. Roy}
\email{smroy@hbcse.tifr.res.in} \affiliation{HBCSE,Tata Institute of Fundamental Research, Mumbai}

{\begin{abstract}

 Monitoring photon quadratures and free masses are useful tools to detect small disturbances such 
 as gravitational waves.Here we report a large class of states 
 for  photon quadratures  and free masses potentially useful for this purpose: (1)''generic coherent states'' (GCS) of photons , 
 whose width is independent of time and uncertainty product $\sigma (x) \sigma (p) $ is arbitrarily large 
 (a generalization of the minimum uncertainty Schr\"odinger coherent 
 states \cite{Schrodinger} ) and (2) ``squeezed generic contractive states'' (SGCS) for photons and free masses 
 (a generalization of the Yuen states \cite{Yuen}) 
 whose width decreases with time ,uncertainty product is arbitrarily large, and the covariance squared $ <\{\Delta \hat{x}, \Delta \hat{p} \}>^2 $
 has an arbitrary value within the allowed range $(0,4 \sigma ^2 (x) \sigma ^2 (p) -1\>)$.\\
 ---------------------Dedicated to the 125th birth anniversary of S. N. Bose.
\end{abstract}


\maketitle

% May skip begin and end figure options for lectures
% Caption works in the (begin and end) figure environment only
%\begin{figure}[ht]
%\begin{center}
% Use LaTeX and DVItoPDF while using EPS files
% \includegraphics[width=.75\columnwidth]{position_density.eps}
% Use PDFLaTeX while using JPG, PNG, PDF image files
%\includegraphics[width=.75\columnwidth]{W3.jpg}
%\caption{Position density}
%\label{fig:pos_dens}
%\end{center}
%\end{figure}
%\begin{center}
 %\includegraphics[width=.75\columnwidth]{neumann.jpg}
%\end{center}

 {\bf History}. S. N. Bose's 1924 paper ``Planck's Law and Hypothesis of Light Quanta '' founded quantum statistics even before 
 quantum mechanics was born. Naturally, it is one of the pillars of quantum optics . Here we construct quantum states of 
 optical quadratures which are non-spreading and hence useful for accurate monitoring of small disturbances.
 
 Much before the actual discovery of gravitational waves \cite{Abbott} it was realised that accurate monitoring of 
position of an oscillator and of a free mass, including intrinsic quantum uncertainties, are important for 
gravitational wave interferometers \cite{Thorne}. Monitoring accuracy is significantly restricted  due to the 
nearly ubiquitous ``spreading of wave packets '' suggested by the heuristic standard quantum limit 
(SQL) (\cite{Braginsky},\cite{Caves1980}).Fortunately, Yuen \cite{Yuen}
 discovered that there are contractive states of free masses for which the SQL is incorrect.Recently one of us (SMR) \cite{SMR2018} 
 has obtained rigorous quantum limits (RQL) on monitoring free masses ,oscillators and photon quadratures ,and the 
 corresponding maximally contractive states. Consistent with these RQL we present a large class of generic coherent states 
 and generic contractive states likely to be useful for accurate quantum monitoring.
 
 In 1926, referring to the general property of spreading of wave packets, H. A. Lorentz \cite{Lorentz} said ,
 in a letter to Schr\"odinger, 
 ''because of this unavoidable blurring a wave packet does not seem to me to be very suitable for representing things to which 
 we want to ascribe a rather permanent individual existence ''.In his reply  \cite{Schrodinger} constructed the 
 now famous  oscillator  coherent state whose wave packet has a width (and shape) independent of time. 
 At the 1927 Solvay conference Einstein used wave packet spreading to discuss the example of a particle passing through a narrow hole on to a
hemispherical fluorescent screen which records the arrival of the
particle (Fig. 1).

\begin{figure}[ht]
\begin{center}
 \includegraphics[width=.3\columnwidth]{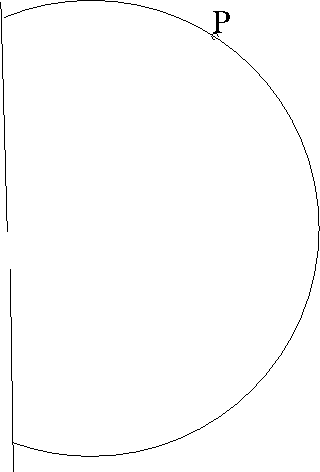}
 \caption {Einstein's single slit example}
\end{center}
\end{figure}

 Suppose that a scintillation is seen at a point $P$ at time
$t = T$, and suppose that the hole is so narrow that the wave packet
corresponding to the particle is uniformly spread all over the screen
at $t$ slightly less than $T$.  Was the particle somewhere near $P$ at
$t = T - \epsilon$ ($\epsilon$ small)?  Ordinary quantum mechanics
says that the probabilities at $t = T - \epsilon$ for the particle
being found anywhere on the screen are uniform (and not particularly large
in the vicinity of $P$).  Thus the naive history corresponding to the
reality of  particle positions at each time  is
absent. Born's rule says that the wave function gives probabilities 
of a particle being found somewhere if a measurement is made, and not of 
being somewhere.

 The role of wave packet spreading in discussions of quantum foundations was further 
 bolstered by heuristic arguments proposing 
that the accuracy of monitoring position of a free mass $m$ is limited by the standard quantum limit 
(SQL) (\cite{Braginsky},\cite{Caves1980}):

\begin{eqnarray} \label{SQL}
 &\sigma^2 (X(t)) \geq  \sigma^2 (X(0)) +(t^2/m^2) \sigma^2 (P(0)) \\
 &\geq  2(t/m)\sigma(X(0))\sigma (P(0)) \geq \hbar t/m\>,
\end{eqnarray}
where $\sigma^2 (X(t))$ and $\sigma^2 (P(t))$ denote variances of the Heisenberg representation position and momentum operators 
at time $t$.

 For the free mass , the inequality (\ref{SQL}) is actually an equality for Gaussian states,

\begin{eqnarray}
 <p| {\psi (t)}> = (\pi \alpha )^{-1/4} exp [-\frac {(p-\beta )^2 } {2\alpha } -it\frac{p^2 } {2m } ],\nonumber \\
 \sigma ^2 ( P (t)) = \frac{\alpha}{2}, \> \sigma^2 (X(t))=\hbar ^2 \frac{1+ (\alpha t/(m\hbar))^2 } {2\alpha },
\end{eqnarray}
 
% \begin{eqnarray}
% && <x| {\psi (t)}> = \frac {(\alpha /\pi)^{1/4} } {\sqrt{1+i t\alpha /(m\hbar ) } } exp (f),\nonumber\\
% && f=-\frac{\alpha}{2 \hbar^2 \big(1+ t^2 (\alpha /(m\hbar ))^2 \big) } \times\nonumber\\
% && \big[(x-\frac{\beta t}{m})^2-i t\frac{\alpha }{m\hbar}(x^2-(\hbar \beta/\alpha )^2)-2i\hbar \beta x/\alpha   \big]
% \end{eqnarray}

%\begin{eqnarray}
% <x||{\psi (t)}> = \frac {(\alpha /\pi)^{1/4} } {\sqrt{1+i t\alpha /(m\hbar ) } } exp (f),\nonumber\\
 % f=-\frac{\alpha}{2 \hbar^2 \big(1+ t^2 (\alpha /(m\hbar ))^2 \big) } \times\nonumber\\
  %\big[(x-\frac{\beta t}{m})^2-i t\frac{\alpha }{m\hbar}(x^2-(\hbar \beta/\alpha )^2)-2i\hbar \beta x/\alpha   \big]
 %\end{eqnarray}

 Equation ( \ref{SQL}) forms the basis of most discussions of spreading of wave packets. The non-causality of quantum mechanics 
 is in sharp focus, because the same initial state is equally likely to result in clicks at widely separated points on the screen.
 Actually a causal ``hidden variable'' theory which yields the same probability densities of position and 
 momentum as ordinary quantum mechanics exists \cite{Roy-Singh1995}. Nevertheless ``non-classicality'' of the trajectories is 
 inevitable.
 
 Both expanding and contracting wave packets represent departures from classicality or coherence represented by 
 wave-packets of constant width. We shall see that the uncertainty principle limits both the possible rates of 
 expansion and of contraction.
 
{\bf Rigorous quantum limits on contractive and expanding states for a free mass}. Surprisingly, for free masses, Yuen discovered in 
1983 \cite{Yuen} a class of states  
 called 'twisted coherent states' which are 'contractive states', i.e. states whose position uncertainty decreases with time 
 for a certain duration. The  SQL is incorrect for these states. One of us (SMR)\cite{SMR2018} obtained rigorous quantum limits (RQL) valid 
 for all states including contractive states.
 
 For any observable with Schr\"odinger operator $A$ (e.g. position $A=X$ or momentum $A=P$), and any Hamiltonian $H$, 
the Heisenberg operator $A(t)$ at time t and its variance $\sigma^2 (A(t)) $ are defined by,
\begin{eqnarray}
&& A(t)\equiv exp(iHt/\hbar)\> A \>exp (-iHt/\hbar),\> \\
&& \sigma^2 (A(t)) \equiv <{\psi(0)}|(\Delta A(t))^2 | {\psi(0)}>,\\
&& \Delta A(t)\equiv A(t)-<A(t)>,\\
&& <A(t)>\equiv<{\psi(0)}|A(t)|{\psi(0)}>
\end{eqnarray}
where $| {\psi(0)}>$ is the initial state.
 
 For a free mass, $H=P^2/(2m)$ ; the Heisenberg equation yields,
 \begin{equation}
  \Delta X(t)= \Delta X(0) +(t/m)\Delta P(0),
 \end{equation}
 and hence,
\begin{eqnarray}\label{Heisenberg}
 \sigma^2 (X(t))=\sigma^2 (X(0))+ (t^2/m^2) \sigma^2 (P(0)) + \nonumber \\
 (t/m)<{\psi(0)}| \{\Delta X(0), \Delta P(0)\}  |{\psi(0)}>.
\end{eqnarray}
One obtains the SQL (\cite{Braginsky},\cite{Caves1980}) Eq. (\ref{SQL}) if one assumes that the third term on the right-hand side, viz. the 
covariance $<\{\Delta X(0), \Delta P(0)\} >$ is non-negative. Yuen showed that the covariance is in fact negative for certain states.
Nevertheless,  rigorous quantum limits (RQL) can be obtained on the covariance, and hence on $\sigma^2 (X(t)) $.

% \begin{eqnarray}
% && <x| {\psi (t)}> = \frac {(\alpha /\pi)^{1/4} } {\sqrt{1+i t\alpha /(m\hbar ) } } exp (f),\nonumber\\
% && f=-\frac{\alpha}{2 \hbar^2 \big(1+ t^2 (\alpha /(m\hbar ))^2 \big) } \times\nonumber\\
% && \big[(x-\frac{\beta t}{m})^2-i t\frac{\alpha }{m\hbar}(x^2-(\hbar \beta/\alpha )^2)-2i\hbar \beta x/\alpha   \big]
% \end{eqnarray}

%\begin{eqnarray}
 %<x||{\psi (t)}> = \frac {(\alpha /\pi)^{1/4} } {\sqrt{1+i t\alpha /(m\hbar ) } } exp (f),\nonumber\\
  %f=-\frac{\alpha}{2 \hbar^2 \big(1+ t^2 (\alpha /(m\hbar ))^2 \big) } \times\nonumber\\
  %\big[(x-\frac{\beta t}{m})^2-i t\frac{\alpha }{m\hbar}(x^2-(\hbar \beta/\alpha )^2)-2i\hbar \beta x/\alpha   \big]
 %\end{eqnarray}

Using $ [\Delta X(0), \Delta P(0)]=i\hbar $ , we have,
\begin{eqnarray}
 &<{\psi(0)}| \{\Delta X(0), \Delta P(0) \}  |{\psi(0)}>+i\hbar \nonumber\\
 &=2 <{\psi(0)}| \Delta X(0) \Delta P(0)  |{\psi(0)}>.
\end{eqnarray}
Cauchy-Schwarz inequality on the right-hand side yields,
\begin{eqnarray}\label{Cauchy}
 &\big(<{\psi(0)} |\Delta X(0) \Delta P(0) +\Delta P(0) \Delta X(0) |{\psi(0)}>\big)^2\nonumber\\
 &\leq 4 \sigma^2(X(0)) \sigma^2(P(0)) -\hbar ^2  ,
 \end{eqnarray}
 which is a rearrangement of the usual Schr\"odinger-Robertson uncertainty relation \cite{Kennard}.
 
 Substituting this into Eq. (\ref{Heisenberg}) we have the rigorous quantum limits (RQL) \cite{SMR2018} 
 on expansion and contraction of wave packets,
 \begin{eqnarray}\label{RQL}
 && |\sigma^2(X(t))-\sigma^2(X(0)) -(t/m)^2 \sigma^2(P(0))|\nonumber \\
&& \leq (t/m) \sqrt{4 \sigma^2(X(0))\sigma^2(P(0)) -\hbar^2 }.
\end{eqnarray}
{\bf valid for arbitrary states}.
The only states saturating the inequalities are those which obey 
\begin{eqnarray}\label{MC state}
&& \Delta P(0)  |{\psi(0)}> =i\lambda \Delta X(0)  |{\psi(0)}> ,\\ 
&& <X' |{\psi (0)}>= \big ( \frac{Re \lambda } {\pi \hbar } \big )^{1/4} \nonumber \\
&&\times exp \big(\frac{i<P(0)> X'} {\hbar } -\frac{\lambda (X'-<X(0)>)^2 } {2\hbar }\big ),\label{optimal state}
\end{eqnarray}
with $ Re \lambda > 0$, 
\begin{eqnarray}
 |Im \lambda | =  \frac{1}{2 \sigma^2(X(0))} \sqrt{4 \sigma^2(X(0))\sigma^2(P(0)) -\hbar^2 } ,\nonumber\\
 \sigma^2(X(0)) =\hbar/(2  Re \lambda) ,\>\sigma^2(P(0)) =\hbar | \lambda |^2 /(2  Re \lambda) ,
\end{eqnarray}

and,
\begin{eqnarray}
 <{\psi(0)}| \{\Delta X(0) ,\Delta P(0) \}  |{\psi(0)} >\nonumber\\
= \mp  \sqrt{4 \sigma^2(X(0))\sigma^2(P(0)) -\hbar^2 }, \>if \> Im \lambda = \pm |Im \lambda |
\end{eqnarray}

The  positive sign of  $Im \lambda $ corresponds to {\bf maximally contractive} 
(essentially Yuen states \cite{Yuen} ),and the negative sign of $Im \lambda $ to  
{\bf maximally expanding } wave packets. 

\begin{figure}[ht]
\begin{center}
 \includegraphics[width=.85\columnwidth]{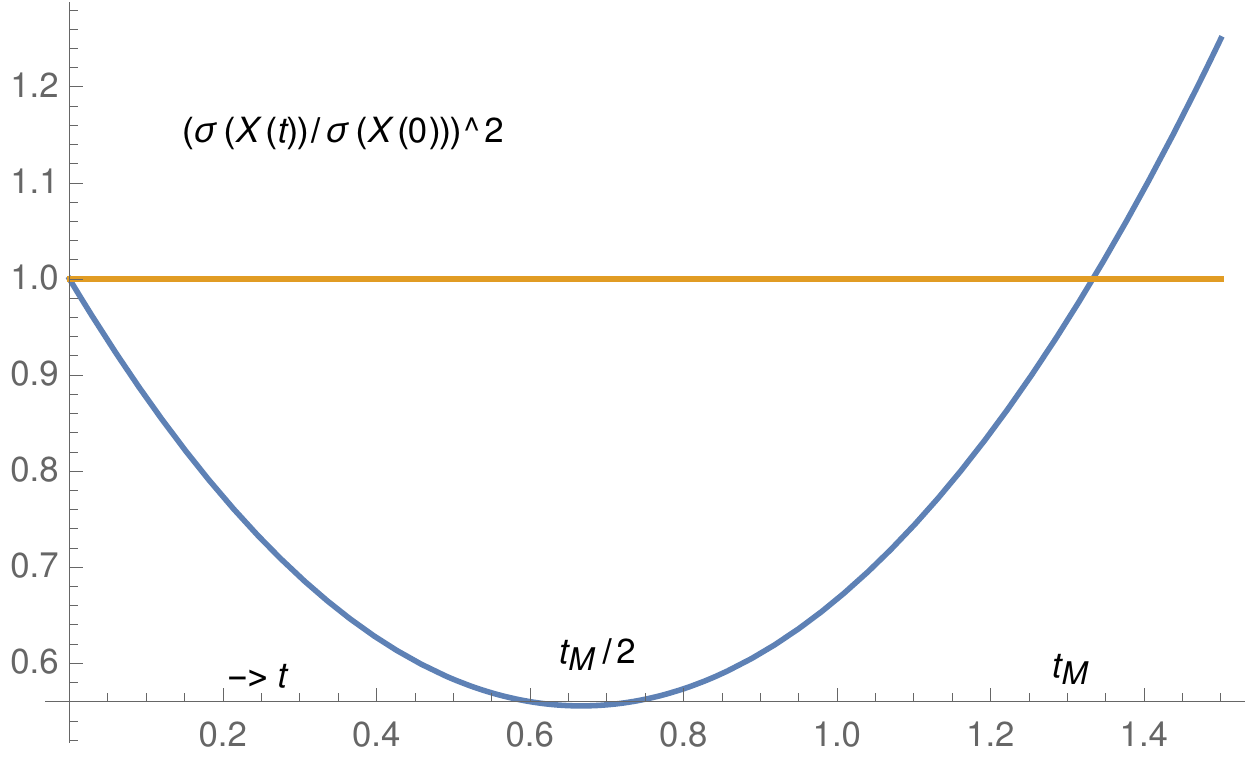}
 \caption{Contractive State}
\end{center}
\end{figure}

Fig. 2 shows that for the initial state (\ref{MC state} ) with positive $Im \lambda $ , the state at time $t$ remains contractive upto  $t=t_M/2$, 
where,
\begin{equation} \label{tM1}
 t_M= \frac{m}{ \sigma^2(P(0)) } \sqrt{4 \sigma^2(X(0))\sigma^2(P(0)) -\hbar^2 } ,
\end{equation}
%It is useful to rewrite the left-hand side of the inequality (\ref{RQL}) in two alternative forms:
%\begin{eqnarray}
%&& \sigma^2(X(t)) \ge \big(\frac{\hbar}{2\sigma (P(0))}\big)^2 +\big(\frac{\sigma (P(0))}{m} \big)^2 (t-\frac{1}{2}t_M)^2 \label{RQL1}\\
%&&=\frac{t}{m} \big(2\sigma(X(0))\sigma(P(0))-\sqrt{4 \sigma^2(X(0))\sigma^2(P(0)) -\hbar^2 } \big)\nonumber\\
%&&+\big( t\frac{\sigma (P(0))}{m}-\sigma(X(0)\big)^2 , \label{RQL2}
%\end{eqnarray}

and, for a given uncertainty product, by choosing $(t/m) \sigma ^2 (P(0))=\sigma(X(0))\sigma (P(0))$, 
$\sigma^2(X(t))$ can be made $\approx t\hbar ^2 /(4m \sigma(X(0))\sigma(P(0)) )$ for a large uncertainty product, and can be much smaller than the heuristic standard 
 quantum limit $\hbar t/m $ .

%However, for given $\sigma(X(0))$ and twist parameter, a class of Generalized Twisted Coherent States 
%(GTCS) \cite{B } can be defined which remain contractive for longer durations than the TCS.

{\bf Rigorous Quantum Limits on Monitoring Photon Quadratures . }

The single mode photon Hamiltonian is,
\begin{equation}
 H= \hbar \omega \> a^\dagger a =\frac{1}{2}\hbar \omega (p^2+x^2-1) ,
\end{equation}
where the quadrature operators $x,p$ are given by 
\begin{equation}
  a=(x+ip )/\sqrt{2} ,\> a^\dagger=(x-ip)/\sqrt{2}.
\end{equation}
The Heisenberg equations of motion yield,
\begin{eqnarray}\label{sigmax(t)}
 &&\sigma^2 (x(t))=\cos ^2(\omega t)\sigma^2 (x(0))+  \sin ^2(\omega t)\sigma^2 (p(0)) \nonumber \\
&&+ \frac{1}{2} \sin (2\omega t)<{\psi(0)}| \{\Delta x(0), \Delta p(0)\}  |{\psi(0)}>.
\end{eqnarray}
As before,  $[\Delta x(0) ,\Delta p(0)]=i $ and  the Schr\"odinger-Robertson uncertainty relations yield 
the RQL \cite{SMR2018},
 
\begin{eqnarray}\label{RQLOSC1}
&&  |\sigma^2 (x(t)) -\cos ^2(\omega t)\sigma^2 (x(0)) -\sin ^2(\omega t)\sigma^2 (p(0))|\nonumber\\
&& \leq \frac{1}{2} |\sin (2\omega t)| \sqrt{4 \sigma^2(x(0)) \sigma^2(p(0)) -1 }
\end{eqnarray}
which corresponds to Eqn.(\ref{RQL} ) for a free mass. 
The extremal states saturating these RQL are complex Gaussians corresponding to  to Equations (\ref{MC state}), (\ref{optimal state})  ; both  the  
maximally contractive (MCON) and maximally expanding (MEXP) states can be designated as  
 'twisted coherent states' \cite{Yuen} or 'squeezed coherent states' (SCS) ,
\begin{eqnarray}
&&SCS: (b-\beta)|{\psi (0)}>=0, \>with \> b = \mu a + \nu a^\dagger, \\ \label{SCSb}
&&\alpha \equiv < \psi (0)|a |{\psi (0)}>, \>\beta= \mu \alpha + \nu \alpha ^*, \nonumber\\
&& \mu = \cosh r , \> \nu =  e^ {i\theta} \sinh r, \\ \label{TCS1} 
&&|{\psi(0) }>=| {\alpha,r \exp {(i\theta ) } }>\equiv D(\alpha,a) S(\xi) | {0}> ,\label{SC}
\end{eqnarray}
where $r>0$ is the squeezing parameter , $\theta$ is real and $|0>$ denotes 
the vacuum state ;here the unitary displacement operator $ D(\alpha,a)$ and squeeze operator $S(\xi)$ are,
\begin{eqnarray}
 && D(\beta,b)=D(\alpha,a)=\exp {(\alpha a^\dagger -\alpha^* a) } , \nonumber \\
 && S(\xi)=\exp {\frac{1}{2} \big (\xi^*a^2-\xi a^{\dagger 2} \big) },\> \xi \equiv r \exp {(i\theta ) }.
\end{eqnarray}
They obey ,
\begin{eqnarray}
&& D^\dagger (\alpha,a) a D (\alpha,a) = a+ \alpha, \nonumber\\
&& D^\dagger (\beta,b) b D (\beta,b) = b + \beta ,\nonumber\\
&& S(\xi) a S^\dagger(\xi) =  a \cosh r + a^\dagger  e^ {i\theta} \sinh r=b \>.
\end{eqnarray}
Explicit values for the standard deviations and covariance in the SCS (\ref{SC}) are then easily derived,

\begin{eqnarray}
 &&\sigma^2 (x(0))=1/2 \big(cosh (2r)-cos (\theta) sinh (2r) \big)\nonumber\\
 &&\sigma^2 (p(0))=1/2 \big(cosh (2r)+cos (\theta) sinh (2r) \big)\nonumber\\
 &&<{\psi(0)}| \{\Delta x(0), \Delta p(0)\}  |{\psi(0)}>=-sin (\theta) sinh (2r)\nonumber\\
 && =- sgn (sin (\theta) )\sqrt{4 \sigma^2(x(0)) \sigma^2(p(0)) -1 }.\nonumber
\end{eqnarray}
For $r>0$, the state is squeezed ,i.e. $\sigma^2 (x(0)) < \sigma^2 (p(0)) $, if $ cos (\theta) >0$, and the 
state is contractive for small positive $t$ if $ sin (\theta) >0$ .
The squeezed coherent states of negative covariance, being contractive ,  have  been 
utilised in precision measurements with gravitational interferometers \cite{squeezed measurements}.

{\bf Generic Coherent States (GCS) of arbitrarily large uncertainty product}
The Schr\"odinger coherent states $|\alpha >= D(\alpha,a)|0>$ have minimum uncertainty product $\sigma(x(0))\sigma(p(0))=1/2$,
and time-independent $\sigma(x(t)),\sigma(p(t))$.
Roy and Singh \cite{Roy-Singh1982} noted that the property of time-independent width of the wave packets also holds for the 
generalised coherent states ,

\begin{eqnarray}
 &&|\psi(\alpha,n)>=D(\alpha,a)|n>;\>\sigma(x(t))=\sigma(p(t))=\sqrt{n+1/2};\nonumber\\
 && a^\dagger a|n>=n |n>.
\end{eqnarray}

 We show here that the property of time-independent width of the wave packet holds for a  class of states 
 much larger than these displaced oscillator eigen states. We  
 call this new class, ``Generic coherent states'' (GCS); they have arbitrarily large continuous values of the  uncertainty product. 
 
 From the time development 
 equation (\ref{sigmax(t)}), denoting expectation value of an operator $A$ in the initial state $|\psi>$ by $<A>$ ,
 we see that $\sigma (x(t))$ is time-independent if and only if,
 
 \begin{eqnarray}
 && GCS:\> <(\Delta x(0))^2> =  <(\Delta p(0))^2> , \>and\>\nonumber\\
 && <\{\Delta(x(0)),\Delta(p(0)) \}>=0.\>
 \end{eqnarray}
Using,
\begin{eqnarray}
 && (\Delta x(0))^2 - (\Delta p(0))^2+i\{\Delta(x(0)),\Delta(p(0)) \}\nonumber\\
 && =2 (\Delta a )^2,
\end{eqnarray}

the GCS conditions  are equivalent to,
\begin{equation}
 GCS\>condition\>:\> <\psi|(\Delta a )^2|\psi>=0\>.
\end{equation}
The GCS include the usual coherent states $\Delta a |\psi>=0 $ as a special case.

We now have the theorem:

If $|\phi>$ is a normalized state obeying 
\begin{equation}\label{generic coherence}
 <\phi|a|\phi>= 0,\> and \> <\phi|a^2|\phi>= 0,
\end{equation}

and 
\begin{equation}
 |\psi (\alpha,\phi)>\equiv D(\alpha,a) |\phi> ,
\end{equation}
where $\alpha$ is an arbitrary complex parameter, then
$|\psi (\alpha,\phi)> $ is a generic coherent state (GCS).

For proof it suffices to note that 
\begin{eqnarray}
 &&<\psi(\alpha,\phi)|a-\alpha|\psi(\alpha,\phi)>\nonumber\\
 && = <\phi|D^\dagger (\alpha,a)\>(a-\alpha)\>D(\alpha,a) |\phi>\nonumber\\
 && =<\phi|a|\phi>=0, \nonumber
 \end{eqnarray}
 and
 \begin{eqnarray}
 &&<\psi(\alpha,\phi)|(a-\alpha)^2|\psi(\alpha,\phi)>\nonumber\\ 
 && = <\phi|D^\dagger (\alpha,a)\>(a-\alpha)^2\>D(\alpha,a) |\phi>\nonumber\\
 && =<\phi|a^2|\phi>=0.\nonumber
\end{eqnarray}

When $|\phi>=|n>$, we get the Roy-Singh \cite{Roy-Singh1982} generalised coherent states; but the possible states $|\phi>$ 
form a much larger set allowing arbitrarily large continuous values of the uncertainty product :
\begin{eqnarray}\label{continuous uncertainty}
 &&\sigma^2 (x(0))=\sigma^2 (p(0))=\sigma (x(0)) \sigma (p(0))=\bar{n}+1/2 ;\nonumber\\
 &&<\psi(\alpha,\phi)|\{\Delta x(0), \Delta p(0)\}|\psi(\alpha,\phi) >=0;\nonumber\\
 && \bar{n} \equiv <\phi|a^\dagger a|\phi>.
 \end{eqnarray}

 It remains only to show that states $|\phi>$ giving arbitrary non-negative values of $\bar{n} $ exist.
Let  
\begin{equation}\label{phi}
 |\phi >= \sum_{m=n}^N c_m |m > \> and <\phi|\phi>=1.
\end{equation}
We assume $N\geq n+3$ and solve Equations (\ref{generic coherence}) to get

\begin{eqnarray}
 && \begin{bmatrix}
   c_{n+1} \sqrt{n+1} & &  c^*_{N-1}\sqrt{N}\\
  \\
    c_{n+2} \sqrt{(n+1)(n+2)} &  & c^*_{N-2} \sqrt{(N-1)N}
 \end{bmatrix}
  \begin{bmatrix}
  c^*_n \\
  \\
  c_N
 \end{bmatrix}\nonumber\\
 && \nonumber\\
&& = -
 \begin{bmatrix}
 \sum_{m=n+1}^{N-2}c^*_m c_{m+1} \sqrt{m+1} \\
 \\
 \sum_{m=n+1}^{N-3}c^*_m c_{m+2} \sqrt{(m+1)(m+2)},
 \end{bmatrix}
 \end{eqnarray}
 where the summations on the right-hand side are to be replaced by zero when the upper limit on $m$ is less than the lower 
 limit. This pair of linear eqns.  can be solved explicitly for $c_n$ and $c_N$, in terms of 
all the other non-zero $c_m$'s . We omit the explicit general solution because  it is elementary.
We just  quote the solution for the  special case $N=n+3$, 
\begin{eqnarray}
&&( |c_{n+1} |^2 -|c_{n+2}  |^2 )
\begin{bmatrix}
 c^*_n \sqrt{n+1} \\
  c_{n+3}\sqrt{n+3}
  \end{bmatrix}\nonumber\\
&&\nonumber\\
&&= - \sqrt{n+2} c^* _{n+1}  c_{n+2}  
  \begin{bmatrix}
   c^* _{n+1}  \\
   -c_{n+2}
  \end{bmatrix}
  ,\nonumber
\end{eqnarray}

It is elementary to check that the vast class of states ( \ref{phi}) obeying (\ref{generic coherence} ) 
can yield any value of $\bar{n} $. E.g. if
\begin{equation}
 |\phi> = \sum_{r=0}^s c_{3r}|3r>\>;and\> \sum_{r=0}^s | c_{3r}|^2=1,\nonumber
\end{equation}
then, equations (\ref{generic coherence} ) are obeyed , and
\begin{equation}
 \bar{n} = \sum_{r=0}^s 3r| c_{3r}|^2 \in [0,3s]\>,
\end{equation}
which can equal any value in the continuous interval $[0,3s]$.Thus the GCS with continuously varying uncertainties  
(\ref{continuous uncertainty}) are obtained.

{\bf Squeezed generic coherent states (SGCS)}.
Using the states $|\phi>$ to replace the vacuum state $|0>$  leads to the class of generic coherent states (GCS) with arbitrarily large 
continuous uncertainty products. We may similarly  generalize the squeezed coherent states (SCS) of maximum possible magnitude of the 
covariance $|<\{\Delta x(0), \Delta p(0) \}>| $ (i.e. maximally contractive or maximally expanding state) to squeezed 
generic coherent states (SGCS) which can have any value of the covariance allowed by the uncertainty principle. 

Consider, the states 
\begin{equation}
 |\psi (\alpha,\xi,\phi)>=D(\alpha,a) S(\xi) |\phi> ,
\end{equation}
which are obtained by replacing $|0>$ in the SCS by the  state  $ |\phi> $ .
These states obey the  SGCS conditions, 
\begin{eqnarray}
&&SGCS:\> <\psi(\alpha,\xi,\phi)|b-\beta|\psi(\alpha,\xi,\phi)>= 0,  \\
&& <\psi(\alpha,\xi,\phi)|(b-\beta)^2|\psi(\alpha,\xi,\phi)>= 0,\\
&& \> i.e. <\Delta b >=0\>,\>and\> < (\Delta b )^2 >=0,
\end{eqnarray}
which are obvious generalizations of the SCS conditions ( \ref{SCSb}).

  Unlike the SCS wave functions, the SGCS wave functions are not complex Gaussians. E.g., when $|\phi>=|n>$ ,
  using $(b^\dagger \> b-n)S(\xi)|n>=0$ we get the displaced and scaled oscillator eigen functions,
  \begin{eqnarray}
  && <x|S(\xi)|n>=\frac{1}{ \sqrt{|\mu-\nu|h_n } }H_n(\frac {x}{|\mu-\nu|} )\>exp (-1/2\lambda x^2),\nonumber\\
  && <x|\psi(\alpha,\xi,\phi)>\nonumber\\
  &&=<x-\sqrt{2}\alpha_1 |S(\xi)|n> \>exp (i\sqrt{2}\alpha_2 (x-\frac{\alpha_1}{\sqrt{2} })),\\
  && h_n=\sqrt{\pi}2^n n!;\>\lambda =\frac{ \mu+\nu} { \mu-\nu};\>\alpha_1=Re \alpha\>;\alpha_2=Im \alpha .
  \end{eqnarray}
For general $|\phi>$ (calculated in the above section), we obtain a generalization of the $SCS$ expressions,
\begin{eqnarray}
 &&SGCS:\>\sigma^2 (x(0))=(\bar{n}+1/2) \big(cosh (2r)-cos (\theta) sinh (2r) \big),\nonumber\\
 &&\sigma^2 (p(0))=(\bar{n}+1/2) \big(cosh (2r)+cos (\theta) sinh (2r) \big),\nonumber\\
 &&<\psi(\alpha,\xi,\phi)| \{\Delta x(0), \Delta p(0)\}  |\psi(\alpha,\xi,\phi)>\nonumber\\
 &&=-(2\bar{n}+1)\>sin (\theta) sinh (2r)\nonumber\\
 && =- sgn (sin (\theta) )\sqrt{4 \sigma^2(x(0)) \sigma^2(p(0)) -(2\bar{n}+1)^2 }.\nonumber
\end{eqnarray}
Time development of these generic contractive or expanding wave packets follows from Eqn. (\ref{sigmax(t)} ) using 
$ |\psi(\alpha,\xi,\phi)>$ as the initial state.

{\bf Overcompleteness of the SGCS}.
Let, $ S(\xi) |\phi> \equiv |\psi(\xi,\phi)> $. Then,$|\psi(\alpha,\xi,\phi)>=D(\alpha,a)|\psi(\xi,\phi)>$. Hence,
\begin{eqnarray}
&& \int d^2 \alpha \frac{1}{\pi}|\psi(\alpha,\xi,\phi)>  \> <\psi(\alpha,\xi,\phi)| \nonumber\\
&& =\int d^2 \alpha \frac{1}{\pi}D(\alpha,a)|\psi(\xi,\phi)>\>< \psi(\xi,\phi)|D^\dagger(\alpha,a).
 \end{eqnarray}
 The integration over $\alpha$ and the fact that $|\psi(\xi,\phi)> $ is a normalized state yields 
 the overcompleteness relation,
 \begin{eqnarray}
  && <x|\int d^2 \alpha \frac{1}{\pi}|\psi(\alpha,\xi,\phi)>  \> <\psi(\alpha,\xi,\phi)|x'> \nonumber\\
&& = \delta (x-x')
 \end{eqnarray}

{\bf Free Mass}. The SGCS for the dimensionless photon variables $x,p$ can also be used as initial states 
for a free mass $m$ ,using $X=x\sqrt{ \hbar/(m\omega)} ,P= p\sqrt{m\hbar \omega }$. We then
 find the time development equation for a free mass ,
\begin{eqnarray}
 \sigma^2 (X(t))=\sigma^2 (X(0))+ (t^2/m^2) \sigma^2 (P(0)) \nonumber \\
 -(\hbar \>t/m)sgn (sin (\theta) )\sqrt{4 \sigma^2(x(0)) \sigma^2(p(0)) -(2\bar{n}+1)^2 }\nonumber.
\end{eqnarray}
The third term on the right-hand side ,where the square root involves the dimensionless $x(0),p(0)$ of the last section ,exhibits all possible 
rates of contraction and expansion of wave packets allowed by the uncertainty principle. 
 
{\bf Position Measurements On Free Masses and Harmonic Oscillators Using Contractive States}.
In order to exploit the new possibilities allowed by the contractive states which violate the SQL (but obey the RQL), 
 the Ozawa measurement model for system-meter interaction \cite{Ozawa1988} which improves on the von Neumann model \cite{von Neumann} has been used .
 The basic idea is to make successive measurements of appropriate duration with meters prepared in identical contractive states such that 
 after each measurement the system is left in the contractive state in which the meter was prepared, and between measurements there is 
 contractive evolution with the system Hamiltonian. Details of this can be found in ( \cite{Ozawa1988} , \cite{SMR2018}) , of 
 continuous measurement methods in \cite{Continuous}, and of actual experimental realizations in
  \cite{squeezed measurements} . The new generic coherent states (GCS) and 
 the  generic contractive states among the squeezed generic coherent states (SGCS) are expected to be useful for measurements necessary 
 for accurate `quantum monitoring'.
 
{\bf Acknowledgements}. One of us (SMR) thanks the Indian National Science Academy for the INSA honorary scientist position at HBCSE, TIFR,
and Dipan Ghosh for the invitation to write this article and several editorial suggestions . Priyanshi Bhasin and Ujan Chakraborty 
thank the NIUS program of HBCSE for making this collaboration possible.


\begin{thebibliography}{99}

 \bibitem{Schrodinger}
 E. Schr\"odinger, reply to Lorentz, reprinted in ``Letters on wave mechanics'' (Philosophical Library, New York, 1967) and 
 Naturwiss. {\bf 14},664 (1926).
 
 \bibitem {Yuen}
H. P. Yuen, ``Contractive States and the Standard
Quantum Limit for Monotoring Free-Mass Posi-
tions'',
Phys. Rev. Letters {\bf 51}, 719 (1983);
R. Lynch, Comment on "Contractive States and
the Standard Quantum Limit for Monitoring
Free-Mass Positions", Phys. Rev. Lett. {\bf 52}, 1729
(1984), and R. Lynch, Phys. Rev. Lett. {\bf 54}, 1599 (1985);
C. M. Caves, Defence of Standard Quantum Limit
for Free-Mass Position, Phys. Rev. Lett. 54, 2465
(1985).
 

 \bibitem{Abbott}
 B. P. Abbott et al (LIGO Scientific Collaboration) and Virgo Collaboration, 
 Phys. Rev. Lett. {\bf 116},061102 (2016).
 \bibitem{Thorne}
   E.g.  K. S. Thorne, R. W. P. Drever,C. M. Caves,  M. Zimmermann, and V. D. Sandberg,
  Phys. Rev. Lett. {\bf 40},667,(1978);R. Weiss, in {\it Sources of Gravitational Radiation}, Editor L. Smarr 
 (Cambridge University Press, Cambridge,1979); V. B. Braginsky,Y.I. Vorontsov and K. S. Thorne,
   Science {\bf 209} (4456)547 (1980); B. Abbott et al (LIGO Scientific Collaboration) 
  New J. Phys.{\bf 11},073032(2009); J. Abadi et al (LIGO Scientific Collaboration) Nature Physics {\bf 7} 962 (2011) 
 : S. L. Danilishin and F. Y. Khalili,
 Living Rev. Relativ. {\bf 15}(1)5 (2012); Y. Ma et al, Nature Physics {\bf 13},776 (2017).
 
 \bibitem{Braginsky}
 V. B. Braginsky and Yu. I. Vorontsov, Ups. Fiz. Nauk {\bf 114},41 (1974) 
 [Sov. Phys. Usp. {\bf 17 },644 (1975) ]
 
 \bibitem{Caves1980}
 C. M. Caves, K. S. Thorne, R. W. P. Drever, V. D. Sandberg, and M. Zimmermann,
 Rev. Mod. Phys. {\bf 52},341(1980).
 
 

\bibitem{SMR2018}
S. M. Roy, 'Rigorous quantum limits on monitoring free masses and harmonic oscillators',
Phys. Rev. A {\bf97}, 032108 (2018).
 
 
\bibitem{Roy-Singh1995} 
S.M. Roy and V. Singh {\it
Mod. Phys. Lett.} \underbar{A10}, 709 (1995) and 
`Deterministic Quantum Mechanics in
One Dimension', p. 434, Proceedings of International Conference on
Non-accelerator Particle Physics, 2-9 January, 1994, Bangalore, Ed. R.
Cowsik (World Scientific, 1995);
L. de Broglie, ``Nonlinear Wave Mechanics, A Causal
Interpretation'', (Elsevier 1960); D. Bohm,  Phys. Rev.
{\bf 85}, 166; 180 (1952); D. Bohm and J.P. Vigier, 
 Phys. Rev. {\bf 96}, 208 (1954). 





 
 \bibitem{Lorentz}
 H. A. Lorentz, in letter to Schr\"odinger, 27 May 1926, reprinted in ``Letters on 
 wave mechanics '', (Philosophical Library, New York, 1967).
 
 
 
 \bibitem{Roy-Singh1982}

 S. M. Roy and V. Singh, ``Generalized coherent
states and the uncertainty principle'', Phys. Rev. D{\bf 25}, 3413 (1982).

 \bibitem{Caves1985}
 C. M. Caves,Phys. Rev. Letters {\bf 54}, 2465 (1985).
 
 
 \bibitem{Ozawa1988}
 M. Ozawa,  Measurement Breaking The Standard
Quantum Limit For Free-Mass Position, Phys. Rev. Letters {\bf 60}, 385 (1988). See also,
 M. Ozawa,Ann. Phys. (N.Y.) {\bf 311}(2),350 (2004), and Current Science 
 {\bf 109}(11),2006 (2015); J. P. Gordon and W. H. Louisell,in {\it Physics of Quantum Electronics}, Ed. P. L. Kelly et al,
p.833 (McGraw-Hill,N.Y.1966 ).

\bibitem{Kennard}
E. Kennard, Zeitschr. Phys. {\bf 44} 326 (1927); H. Weyl, {\it Gruppentheorie und quantenmechanik} (Hirzel, Leipzig, 1928); 
H. Robertson, Phys. Rev. {\bf 34}, 163 (1929) ; E. Schrödinger, "Zum Heisenbergschen Unschärfeprinzip", Sitzungsberichte der 
Preussischen Akademie der Wissenschaften, Physikalisch-mathematische Klasse, 14: 296–303 (1930) [English trans. by A. Agelow 
and M. Batoni,http://arxiv.org/abs/quant-ph/9903100]; H. Robertson, Phys. Rev. {\bf 35}, 667 (1930) .



\bibitem{Arthurs-Kelly}
 E. Arthurs and J. L. Kelly Jr. ,Bell System Tech. J.,{\bf 44}(4),725(1965);
 P. Busch, T. Heinonen and P. Lahti, Physics reports {\bf 452}(6),155((2007);
 P. Busch, P. Lahti, and R. F. Werner,Rev. Mod. Phys. {\bf 86}(4),1261(2014);
 S. M. Roy, Current Science {\bf 109}(11)2029(2015);
 
 \bibitem{squeezed}
  D. Walls, Squeezed states of light, Nature 306, 141 (1983); M. O. Scully and M.S. Zubairy,'Quantum Optics', Cambridge Univ. Press (1997);
   C. Gerry and P. L. Knight , 'Introductory Quantum Optics' Cambridge Univ. Press (2005);
 R. Loudon, 'The Quantum Theory of Light' (Oxford University Press, 2000); D. F. Walls and G.J. Milburn, 'Quantum Optics', Springer Berlin (1994);
 C W Gardiner and Peter Zoller, "Quantum Noise", 3rd ed, Springer Berlin (2004).

 \bibitem{Khalili}
 F. Y. Khalili et al, Phys. Rev. Lett. {\bf 105},070403 (2010).
 
 \bibitem{squeezed measurements}
 Grote, H.; Danzmann, K.; Dooley, K. L.; Schnabel, R.; Slutsky, J.; Vahlbruch, H. (2013). "First Long-Term
Application of Squeezed States of Light in a Gravitational-Wave Observatory" (http://arxiv.org
/abs/1302.2188). Phys. Rev. Lett. 110: 181101. arXiv:1302.2188 (https://arxiv.org/abs/1302.2188) ;
J. Aasi et al, 'Enhancing the sensitivity of the LIGO gravitational detector by using squeezed states of light',
arXiv:1310.0383[uant-ph] ; 
Markus Aspelmeyer, Pierre Meystre, and Keith Schwab,' Quantum optomechanics',
 Phys. Today 65(7), 29 (2012); 
Goda, K. et al. ``A quantum-enhanced prototype
gravitational-wave detector''. Nature Physics 4, 472-476
(2008); Abadie, J. et al. ``A gravitational wave observatory operat-
ing beyond the quantum shot-noise limit''. Nature Physics, 7, 962-965 (2011);
 Harry, G.M. et al. ``Advanced LIGO: the next generation
of gravitational wave detectors''. Classical Quant. Grav. 27, 084006 (2010).

\bibitem{von Neumann}
 J. von Neumann, Chap. 6 , ' Mathematical Foundations of Quantum
 Mechanics' (Princeton University, Princeton, New Jersey, (1955).

\bibitem{Continuous}
C. M. Caves and G. J. Milburn, Phys. Rev. A 36, 5543
(1987);M. T. Jaekel and S. Reynaud, Europhys. Lett. 13, 301
(1990);G. J. Milburn, K. Jacobs, and D. F. Walls, Phys. Rev. A
50, 5256 (1994); H. Mabuchi, arXiv:quant-ph/9801039v3 (1998).


\end{thebibliography}
\end{document}